\documentclass{iaus}
\usepackage{graphicx}

\title[3:1 MMR in 55 Cancri] 
{Formation and transformation of the 3:1 mean-motion resonance in 55
Cancri System}

\author[Zhou, Ferraz-Mello \& Sun]   
{Li-Yong Zhou$^1$, Sylvio Ferraz-Mello$^2$ \and Yi-Sui Sun$^1$}

\affiliation{$^1$ Department of Astronomy, Nanjing University \\
Nanjing 210093, China \\
email: {\tt zhouly@nju.edu.cn; sunys@nju.edu.cn}\\[\affilskip]
$^2$Instituto de Astron\^omico, Geof\'isica, e Ci\^encias
Atmosf\'ericas, Universidade de S\~ao
Paulo \\
Rua do Mat\~ao 1226, 05508-900 S\~ao Paulo, Brazil \\
email: {\tt sylvio@astro.iag.usp.br}}

\pubyear{2008}
\volume{249}  
\pagerange{119--126}
 \jname{Exoplanets: Detection, Formation and
Dynamics} \editors{Y.-S. Sun, S. Ferraz-Mello \& J.-L. Zhou, eds.}
\begin{document}

\maketitle

\begin{abstract}
We report in this paper the numerical simulations of the capture
into the 3:1 mean-motion resonance between the planet b and c in the
55 Cancri system. The results show that this resonance can be
obtained by a differential planetary migration. The moderate initial
eccentricities, relatively slower migration and suitable
eccentricity damping rate increase significantly the probability of
being trapped in this resonance. Otherwise, the system crosses the
3:1 commensurability avoiding resonance capture, to be eventually
captured into a 2:1 resonance or some other higher-order resonances.
After the resonance capture, the system could jump from one orbital
configuration to another one if the migration continues, making a
large region of the configuration space accessible for a resonance
system. These investigations help us understand the diversity of
resonance configurations and put some constrains on the early
dynamical evolution of orbits in the extra-solar planetary systems.

\keywords{Planetary systems, celestial mechanics, methods:
numerical, methods: analytical}
\end{abstract}

\firstsection 
\section{Introduction}
Up to date, more than 200 extra-solar planetary systems have been
found. Those hosting more than one planet are multiple planet
systems. In some of the multiple planet systems, planets are
observed to be locked in mean-motion resonance (MMR), for example,
the well-known 2:1 MMR in GJ876 system (e.g. \cite[Marcy et al.
2001]{Marcy01}) and the 3:1 MMR in 55 Cancri system (e.g. \cite[Zhou
et al. 2004]{Zhou04}). In a 3:1 MMR, at least one of the three
resonant angles ($\theta_1=\lambda_1-3\lambda_2+2\varpi_1$,
$\theta_2=\lambda_1-3\lambda_2+2\varpi_2$,
$\theta_3=\lambda_1-3\lambda_2+\varpi_1+\varpi_2$) librates, where
$\lambda_{1,2}$ and $\varpi_{1,2}$ are the mean longitudes and
periastron longitudes of the inner and outer planet respectively.
The resonant systems are particularly attractive because not only of
the complicated dynamics of the resonance but also of interesting
information about its origin and early evolution buried in the
configuration and dynamics.

The 55 Cancri system is the only example of 3:1 MMR found in the
extra-solar planetary systems till now. Four planets have been
reported in this system and two of them, 55 Cnc b and 55 Cnc c, seem
to be locked in a 3:1 MMR. In our previous work, we have found that
they are most likely in one of the three possible configurations
(\cite[Zhou et al. 2004]{Zhou04}). G.Marcy declared in his lecture
in this Symposium that a fifth planet has been dug out from
observing data of this system, enriching the whole story with more
connotations. In the dynamical simulations of their 5-planet
solution (\cite[Fischer et al. 2007]{Fischer07}), they did not found
the resonance between the two planets, although their orbital
periods are very close to the 3:1 commensurability. However, the
current orbit determinations may not be robust enough to determine
the resonant angles, especially for orbits with low eccentricities
as be given in this new determination, thus the resonant character
of this pair of orbits may change as more observations are included
(\cite[Beaug\'e 2008]{Beau08}). Furthermore, what we will discuss in
this paper has some general significance so that we may still adopt
the orbital elements and planetary masses from the previous
literature (\cite[McArthur et al. 2004]{McArthur04}), as listed in
Table \ref{tab1}

\begin{table}
\begin{center}
\caption{Orbital elements and masses. The mass of the central star
is $1.03 M_{\odot}$ (\cite[McArthur et al. 2004]{McArthur04}). In
this paper we adopt the planet masses by assuming $\sin i=1$.}
\label{tab1}
 \scriptsize{
\begin{tabular}{lrrrr}
\hline Parameter & Cancri e & Cancri b & Cancri c & Cancri d \\
 \hline
$M\sin i$ $(M_J)$ & 0.045 & 0.784  & 0.217  & 3.92 \\
$P$ (days) & 2.81 & 14.67 & 43.93 & 4517.4 \\
$a$ (AU) & 0.038 & 0.115 & 0.240 & 5.257\\
$e$  & 0.174 & 0.0197 & 0.44  & 0.327\\
\hline
\end{tabular}}
\end{center}
\end{table}

It is widely accepted that a planet may experience (generally
inward) migration, and through differential migrations of planets,
different commensurabilities among planets can be attained (e.g.
\cite[Nelson \& Papaloizou 2002]{Nelson02}). For example, the 2:1
MMR in GJ876, whose configuration can be obtained through the inward
migration of the outer planet (\cite[Lee \& Peale 2002]{Lee02}).

On the other hand, the resonant capture probability depends on the
order of the resonance, the migration rate and the initial planetary
eccentricity (\cite[Quillen 2006]{Quillen06}). The formation process
of the 3:1 MMR may be very different from the one of the 2:1 MMR. As
the only example of 3:1 MMR in extra-solar planetary systems, the 55
Cancri system deserves a detailed investigation. In this paper, we
will report our numerical simulations of the capture into resonance,
and the evolution of the resonance thereafter, provided the
migration continues.

\section{Numerical simulations of resonance formation}
There are more planets than the two (planet b and c) in our focus of
attention in this planetary system. However, besides these two
planets, the massive one (55 Cnc d) is too far and the close one (55
Cnc e) is too small. Moreover, even the newly reported planet (55
Cnc f) is neither very massive ($M_f\sin i=0.144M_{\rm Jup}$) nor
very close to this planet pair ($a_f=0.781$\,AU, $P_f=260$\,days).
Therefore as a simplified model, it is reasonable to discuss only
the two planets under the influences of the central star and the
disc.

\subsection{Numerical Model}
As usual, the influence of the disc on the planets is simply
simulated by an artificial force, which is acting on the outer
planet to drive it to migrate inward. The force is defined so that
the semi-major axis of the planetary orbit will change following an
exponential law:
 \begin{equation}
 a(t)=a_0+\Delta a\times e^{-t/\tau}.
 \end{equation}
\noindent where $a_0$ is the initial semi-major axis of the planet,
and $\tau$ is the timescale of the migration. The two planets are
assumed to be located initially on orbits with semi-major axes of
1.25\,AU and 0.5\,AU, with an initial semi-major axes ratio of 2.5
(periods ratio 3.95), which will evolve downward to the value
corresponding to the 3:1 MMR (2.08). Obviously the value of $\tau$
determines the migration speed. The smaller the $\tau$, the faster
the migration, and \textit{vice versa}. Different $\tau$ values were
adopted in different papers. The standard migration scenario (type
II) gives an estimate of migration rate (\cite[Ward 1997]{Ward97}):
 \begin{equation}
 \left|\frac{\dot a}{a}\right|=9.4\times10^5\left(\frac{\alpha}{4\times10^{-3}}\right)\left(\frac{H/a}{0.05}\right)^2P^{-1}.
 \end{equation}
Taking the typical parameters $\alpha$ and $H/a$, the $\tau$ for a
planet starting from 1.25\,AU is $\sim 1.43\times10^4$\,yrs. Some
other rules have been applied in different papers. According to
these rules, the $\tau$ would be, for example, $1.56\times10^4$\,yrs
(\cite[Lee \& Peale 2002]{Lee02}), $1.38\times10^4$\,yrs
(\cite[Nelson \& Papaloizou 2001]{Nelson02}), and
$2.65\times10^4$\,yrs (\cite[Kley 2003]{Kley03}; \cite[Kley et al.
2004]{Kley04}). In our simulations,  it is set to be $2\times 10^4,
1\times 10^5, 2\times 10^5$ and $5\times 10^5$\,yrs. Some large
values (meaning slow migration) are adopted here to guarantee the
capture into the 3:1 MMR.

Although it is still not very clear now how a certain disk will damp
or stimulate the eccentricity of an embedded planet, people hope the
disk will circularize an orbit if it was too eccentric. In order to
control the unlikely eccentricity increasing during the artificially
induced migration, we have included an eccentricity-damping force in
our simulations. A parameter $K$ was introduced to describe the
damping rate, with which we control the eccentricity $e$ of a planet
by:
 \begin{equation}
 \left|\dot e/e\right|=K\left|\dot a/a\right|.
 \end{equation}
\noindent In our simulations, the parameter $K$ has the values 0 (no
damping), 1, 10 and 100. The damping is always put on planet 55 Cnc
c with the migrating force modifying its semimajor axis
simultaneously.

In the simulations, the two orbits are nearly coplanar (in the
practice of numerical simulations we adopt an arbitrarily defined
small inclination between these two orbits, say 0.1 degree), and the
initial orbital eccentricities of planet 55 Cnc b ($e_1^0$) and 55
Cnc c ($e_2^0$) are set to be $0.001$ (a typical value for a nearly
circular orbit) as well as $0.01, 0.05, 0.1, 0.2$. For each mode
(with certain $\tau, K, e_1^0, e_2^0$), we numerically integrate 100
systems with randomly selected orbital angles (longitudes of
periastrons, mean longitudes, and ascending nodes). Each testing try
is integrated up to a time of $2\tau$. During an integration, if the
two semi-major axes are locked in a definite value and keep this
value at least for ten percent of the integration time, that is
$\tau/5$, we say the system is in the given commensurability.

\subsection{Results}
Generally, the migration may drive the two planets into a given
commensurability. The inner planet 55 Cnc b will also migrate after
it has been captured by a commensurability. The final configuration
of a system depends sensitively on the migration speed and the
initial conditions. We summarize some remarkable outcomes from our
numerical simulations as follow.

(1) The migration with $\tau\sim 2\times 10^4$\,yrs is too fast to
form the 3:1 MMR. If we start from nearly circular orbits ($e_1^0
=e_2^0 =0.001$) and neglect the eccentricity damping ($K=0$), all
the 200 runs for $\tau= 2\times 10^4$\,yrs and $\tau= 1\times
10^5$\,yrs cross the 3:1 MMR without being trapped and are
eventually captured into the 2:1 resonance. The 3:1 MMR is observed
only when $\tau= 2\times 10^5$\,yrs. When $\tau=2\times 10^4$\,yrs,
the number of simulation runs, in which the 3:1 MMR forms, is very
small, if not null, no matter what $e_1^0, e_2^0$ and $K$ are.

(2) A slow migration favors the formation of the 3:1 MMR. For
initially near circular orbits, when $\tau =2\times10^5$\,yrs, we
have obtained 53 out of the 100 runs trapped in the 3:1 MMR and the
number increases to 100 when $\tau =5\times10^5$\,yrs.

(3) The initial eccentricities strongly affects the resonance
trapping. A nonzero but small initial eccentricity of either orbits
($e_1^0$ or $e_2^0 \sim 0.01)$ will increase significantly the
probability of the 3:1 MMR. For the runs with $e_1^0$ or
$e_2^0=0.01$, although there is no 3:1 MMR when $\tau= 2\times
10^4$\,yrs, we found 163 out of the totally 200 runs trapped in this
resonance when $\tau= 1\times 10^5$\,yrs. Higher initial
eccentricities ($e_1^0, e_2^0 \sim 0.05, 0.10$) will bring in more
higher-order resonances, such as the 5:2, 7:3 and 7:2 MMRs. It is
interesting to see that a few runs were trapped in the 11:4
resonance. Even higher eccentricities ($e_1^0, e_2^0 \sim 0.20$)
generally leads the system to catastrophic planetary scattering.

(4) The eccentricity damping affects the resonance capture too.
Without the eccentricity damping ($K=0$), higher initial
eccentricities would likely lead to high order commensurabilities,
while low initial eccentricities would cause eccentricities to grow
too much after the planets are locked into a low-order resonance. On
the other hand, a high $K$ value restrains the eccentricity
increasing and therefore causes the system to evolve preferably to a
2:1 MMR rather than the 3:1 MMR. For example, when $K=100$ all the
test systems starting from circular orbits ($e_1^0=e_2^0=0.001$) are
driven into the 2:1 MMR if $\tau \leq 2\times10^5$\,yrs. Even in a
very slow migration ($\tau= 5\times10^5$\,yrs), the 2:1 MMR captures
41 out of the 100 runs. So we need a suitable damping rate for the
formation of 3:1 MMR. In all our simulations, the most probable
modes for the formation of the 3:1 MMR were a slow migration
($\tau\geq 1 \times 10^5$\,yrs), moderate initial eccentricities
$(e_{1,2}^0= 0.01\sim0.05)$ and moderate eccentricity damping
($K\sim 10$).

\section{Evolution of orbital configuration in resonance}
After being captured into the 3:1 MMR, the system will continue to
evolve if the migration does not halt. In this section we will
discuss the evolution of the system in the resonance.

\begin{figure}[hc]
\vspace*{-1.0 cm}
\begin{center}
 \includegraphics[width=13cm]{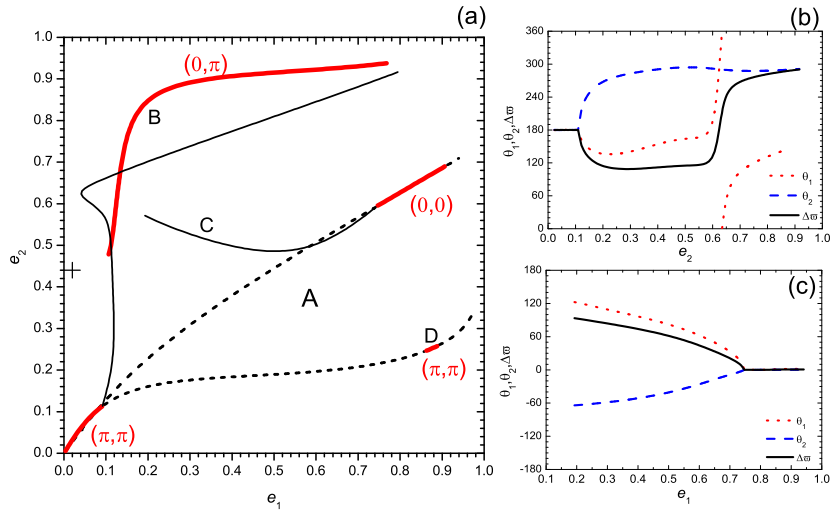}
 \vspace*{-1.5 cm}
 \caption{Periodic solutions in the 3:1 MMR. (a) Periodic solution
 families on the $(e_1,e_2)$ plane. The stable symmetrical ACR solutions
 are thick solid curves, the unstable symmetrical ACR solutions are
 short-dashed curves, and the thin solid curves indicate the asymmetrical
 solutions. The value of $(\theta_1,\Delta\varpi)$ of the symmetrical
 solutions appear as labels along the curves. We also label different solution
 families with \textbf{A}, \textbf{B}, \textbf{C} and  \textbf{D}. The
 eccentricities $(e_1,e_2)=(0.0197,0.44)$ in Table \ref{tab1} are indicated by a cross. (b) The
 variation of the angles $\theta_1$ (dotted), $\theta_2$ (dashed) and
 $\Delta\varpi$ (solid) along the
 solution family  \textbf{A}. Note $\Delta\varpi$ begins from $180^\circ$
 (symmetrical ACR). (c) The same as (b) but for solution family
 \textbf{C}, where $\Delta\varpi$ reaches at symmetrical value $0^\circ$
 at the end. }
   \label{fig1}
\end{center}
\end{figure}

\subsection{Periodic solutions in 3:1 MMR}
Fig.\,\ref{fig1} shows the periodic solutions, both stable and
unstable, symmetrical and asymmetical, computed using the mass ratio
of the planets 55 Cnc b and 55 Cnc c and assuming coplanar orbits.
(For other mass ratios and for details of the method used see
\cite[Michtchenko et al. (2006)]{Micht06}).

In the real system, if we assume that it evolved from nearly
circular orbits through an adiabatical migration, as suggested in
many references, the configuration of two planets captured into a
3:1 resonance should be located on or near the family \textbf{A} of
periodic solutions shown in Fig.\,\ref{fig1}(a). In such a
migration, the angles $(\theta_1, \Delta\varpi)$ evolve from the
symmetrical $(\pi,\pi)$ values to asymmetrical values. Since each
solution family is not connected by stable periodic solutions to
other families\footnote{\textbf{A} and \textbf{B} are not connected
because they have different $\theta_1, \Delta\varpi$ values at the
two crossings in Fig.\,\ref{fig1}(a)}, there is no reason to expect
a system in resonance with orbital configuration far away from the
family \textbf{A}. In fact, the eccentricities $(e_1, e_2)$ in Table
\ref{tab1} are not on the family \textbf{A}. However, we have found
that this configuration could be obtained by a resonance captured
from initial eccentric orbits. An example of such kind of resonance
trapping is shown in Fig.\,\ref{fig2}(a). Surely a perturbation
happening to a system in the equilibrium periodic solution may also
lead to this observed configuration. We just show here another
possibility of the resonance formation.

\subsection{Jump between solution families}
\begin{figure}[h]
 \vspace*{-1.0 cm}
\begin{center}
 \includegraphics[width=14cm]{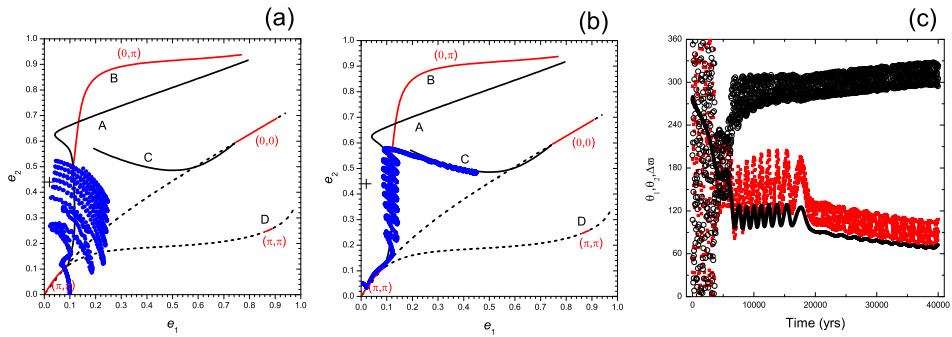}
 \vspace*{-1.5 cm}
 \caption{The evolution of orbital configuration in the 3:1 MMR.
 (a) An example of evolution starting from eccentric initial orbits.
 The eccentricities in Table \ref{tab1} can be reached by the solution in this case. Solution
 families are shown in the same way as in Fig.\,\ref{fig1}, and the
 dots indicate the evolution of one run in our numerical simulations.
 (b) An example of orbital configuration jump from
 solution family \textbf{A} to family \textbf{C}. (c) The time
 variations of angles $\theta_1$ (small dots), $\theta_2$ (open
 circles) and $\Delta\varpi$ (solid thick curve) for the numerical
 simulation shown in (b).}
 \label{fig2}
\end{center}
\end{figure}

If the disc does not disappear just after the resonance capture
happens, the outer planet will continue to migrate inward
accompanied by the inner planet locked in the resonance. During such
after-capture migration stage, the system will evolve along the
solution family \textbf{A} as shown in Fig.\,\ref{fig1}, provided
the migration is ``adiabatic''. But if the migration is faster (not
adiabatic), as some of our numerical simulations, the solution may
jump from a family to another. An example of jump from family
\textbf{A} to \textbf{C} is illustrated in Fig.\,\ref{fig2}(b).
Comparing the variations of angles in Fig.\,\ref{fig2}(c) and
Fig.\,\ref{fig1}(b), (c), it is clear that the jump happens around
$t=1.9\times 10^4$\,yrs. At this moment, if the system goes along
family \textbf{A}, the angles $\theta_1$ and $\Delta\varpi$ will go
up and cross the value $180^\circ$ one after another with the
increasing eccentricity $e_2$. But they fail and turn around
following another evolving route, that is, family \textbf{C}. This
turnaround is due to some perturbation (in our case it's the
migration itself), and it happens around the critical point where
the apsidal difference $\Delta\varpi$ crosses the symmetrical value
$180^\circ$ in an asymmetical ACR. In fact, the stable region of
motion around this critical turning point (on family \textbf{A} with
$(e_1, e_2)= (0.0423, 0.623)$ in Fig.\,\ref{fig1}(a) ), is very
small. Therefore a slight disturbance could drive the system out of
the family \textbf{A}.

The jump from family \textbf{A} to family \textbf{B} has also been
observed in our simulations. The angles $\theta_1 = \theta_2 =
0^\circ$ in family \textbf{B}, that is to say, both $\varpi_1$ and
$\varpi_2$ have to adjust to new values in a jump from \textbf{A} to
\textbf{B}. On the other hand, as we see from
Fig.\,\ref{fig1}(b),(c) and Fig.\,\ref{fig2}(c), we need only an
adjustment of $\varpi_1$ but not $\varpi_2$ for a jump from family
\textbf{A} to \textbf{C}, since this jump does not affect the
$\theta_2$. When the jump happens, the inner planet is on a near
circular orbit ($e_1=0.0423$) while the outer one is on a highly
eccentric orbit ($e_2=0.623$). As a result, it is much easier to
adjust $\varpi_1$ than $\varpi_2$. This analysis tells why we
observe much more \textbf{A} to \textbf{C} jumps than \textbf{A} to
\textbf{B} jumps.

These jumps between different solution families make it possible for
a resonant system to occupy much larger potential volume in the
orbital elements space, increasing the diversity of resonant
configuration in extra-solar planetary systems.

\section{Conclusions}
The 3:1 mean-motion resonance in the 55 Cancri planetary system is
far from a necessary result of the differential planetary migration.
Our numerical simulations showed that the favourable scenario for
the formation of the 3:1 resonance is moderate initial
eccentricities ($0.01\sim 0.05$), relatively slower migration ($\tau
\sim 10^5$\,yrs) and suitable eccentricity damping rate ($K\sim
10$). After being captured in the resonance, the system may exhibit
some evolutions different from the behaviours in an adiabatic
migration, and evolve to some unexpectable orbital configurations.
All these put some constraints on the planetary migration and early
dynamical evolution in the extra-solar planetary systems.

\section*{Acknowledgements}
We thank T.A. Michtchenko for many helpful discussions and for some
of the codes used to find the location of the periodic solutions.
This work was supported by the Natural Science Foundation of China
(No. 10403004), the National Basic Research Program of China (973
Program, 2007CB814800) and Sao Paulo State Research Foundation
(FAPESP, Brazil).


\begin{thebibliography}{}

\bibitem[Beaug\'e (2008)]{Beau08}
{Beaug\'e, C.,} 2008, \textit{Proceedings of IAUS249}, this issue

\bibitem[Fischer \etal\ (2007)]{Fischer07}
{Fischer, D., Marcy, G., Butler, P., Vogt, S., Laughlin, G., Henry,
G., Abouav, D., Peek, K., Wright, J., Johnson, J., McCarthy, C. \&
Isaacson, H.} 2007, \textit{preprint}

\bibitem[Kley (2003)]{Kley03}
{Kley, W.} 2003, \textit{Cel. Mech. \& Dyn. Astron.} 87, 85

\bibitem[Kley, Peitz \& Bryden (2004)]{Kley04}
{Kley, W., Peitz, J. \& Bryden G.} 2004, \textit{A\&A} 414, 735

\bibitem[Lee \& Peale (2002)]{Lee02}
{Lee, M. \& Peale S.} 2002, \textit{ApJ}, 567, 596

\bibitem[Marcy \etal\ (2001)]{Marcy01}
{Marcy, G., Butler, P., Fischer, D., Vogt, S., Lissauer, J., \&
Rivera, E.} 2001, \textit{ApJ}, 556, 296

\bibitem[McArthur \etal\ (2004)]{McArthur04}
{McArthur, B., Endl, M., Cochran, W., Benedict, F., Fischer, D.,
Marcy, G., Butler, P., Naef, D., Mayor, M., Queloz, D., Udry, S. \&
Harrison, T.} 2004, \textit{ApJ}(Letters), 614, L81

\bibitem[Michtchenko, Beaug\'e \& Ferraz-Mello (2006)]{Micht06}
{Michtchenko, T., Beaug\'e, C. \& Ferraz-Mello, S.} 2006,
\textit{Cel. Mech. \& Dyn. Astron.}, 94, 411

\bibitem[Nelson \& Papaloizou (2002)]{Nelson02}
{Nelson, R. \& Papaloizou, J.} 2002, \textit{MNRAS}(Letters), 333
L26

\bibitem[Quillen (2006)]{Quillen06}
{Quillen, A.} 2006, \textit{MNRAS}, 365, 1367

\bibitem[Ward (1997)]{Ward97}
{Ward, W.} 1997, \textit{Icarus}, 126, 261

\bibitem[Zhou \etal\ (2004)]{Zhou04}
{Zhou, L.-Y., Lehto, H., Sun, Y.-S. \& Zheng, J.} 2004,
\textit{MNRAS}, 350, 1495

\end{thebibliography}
\end{document}